\documentclass[preprint,5p,times,twocolumn,fleqn]{elsarticle}

\usepackage{amssymb}
\usepackage{amsmath}
\usepackage{amsfonts}

\usepackage{hyperref}
\hypersetup{colorlinks=true,
    linkcolor=blue,
    filecolor=blue,
   urlcolor=blue,
    citecolor=blue,
}

\journal{Results in Physics}

\begin{document}

\begin{frontmatter}



\title{Optimizing mechanical entanglement using squeezing and parametric amplification}

\author[1]{Muhdin Abdo Wodedo}
\author[2]{Tesfay Gebremariam Tesfahannes \corref{cor}}
\ead{tesfaye.gebremariam@amu.edu.et}
\author[1]{Tewodros Yirgashewa Darge}
\author[3,4]{Berihu Teklu \corref{cor}}
\ead{berihu.gebrehiwot@ku.ac.ae}
\cortext[cor]{Corresponding authors}

\affiliation[1]{organization={School of Natural Sciences},
            addressline={Department of Applied Physics}, 
            city={Adama Science and Technology University},
            postcode={1888}, 
            state={Adama},
            country={Ethiopia}}
 \affiliation[2]{organization={College of Computational and Natural Sciences},
            addressline={Department of Physics}, 
            city={Arba Minch University},
            postcode={21}, 
            state={Arba Minch},
            country={Ethiopia}}
\affiliation[3]{organization={College of Computing and Mathematical Sciences, Department of Applied Mathematics and Sciences, Khalifa University of Science and Technology},
            addressline={127788}, 
            city={Abu Dhabi},
            country={United Arab Emirates.}
            }
\affiliation[4]{organization={ Center for Cyber-Physical Systems (C2PS), Khalifa University of Science and Technology},
            addressline={127788}, 
            city={Abu Dhabi},
            country={United Arab Emirates.}
            }

\begin{abstract}
We propose a scheme of an optomechanical system that optimizes entanglement in nanomechanical resonators through quantum state transfer of intracavity squeezing and squeezed reservoir field sources assisted by radiation pressure. The system is driven by red-detuned laser fields, which enable simultaneous cooling of the mechanical resonators and facilitate the quantum state transfer in a weak coupling {and good cavity limit}. Specifically, the mechanical entanglement is quantified using logarithmic negativity within the bipartite Gaussian states of the two mechanical modes. The results show that several key parameters, including the parametric phase and nonlinear gain of the non-degenerate optical parametric amplifier, the strength of the squeezing reservoir, optomechanical cooperativity,  thermal excitation of phonons, and the temperature of mechanical baths, strongly influence the degree of mechanical entanglement. Hence, the findings indicate that careful tuning of the parameters can enable control over the enhancement of entanglement robustness, suggesting that this optomechanical scheme {provides a viable pathway} for applications in quantum sensing and information processing
\end{abstract}





\end{frontmatter}

\section{Introduction}\label{sec1}

Entanglement is one of the most prominent quantum resources among quantum correlations and has no classical equivalent \cite{horodecki2009quantum, Zhang:24}. Entanglement is experimentally demonstrated \cite{Schori2, Bowen3} and has various essential features in quantum information processing \cite{Braunstein5}. In recent decades, several investigations have shown considerable attention to entanglement in microscopic and mesoscopic systems \cite{bose1999scheme, vedral2004high, deb2008entanglement, Sperling17}. The merits of entanglement encompass quantum communication \cite{ZAK2002, YUAN20101,ADEPOJU2017, TORRESARENAS2019, berihu22}, quantum teleportation \cite{sherson2006quantum,ZUBAREV2022105700,ZHANG2024107632}, quantum cryptography \cite{Bennett92}, quantum metrology \cite{Paris, Teklu_2009,ockeloen2018revealing,PhysRevResearch.2.043338,Teklu_2021,PhysRevResearch.4.033036,berihu23,XIE2023106575, Huang2024}, and quantum computing \cite{shahandeh2019quantum}. 

In recent years, in mesoscopic and macroscopic systems, cavity optomechanics through radiation pressure has become a crucial candidate to capture macroscopic quantum effects with different applications \cite{meystre2013short, aspelmeyer2014cavity}. These quantum effects are highly probable within optomechanical systems via refrigeration of mechanical vibrations \cite{Deng-Gao22}. Recently, passive synchronization in optomechanical resonators coupled through an optical field has been verified in an experimental scheme \cite{WANG2021110717}. Resonance effects in a scheme of optical cavity arrays and one oscillating end mirror enable long-distance optimal optomechanical-entanglement transfer \cite{Shang24}. The Effect of the squeezed-vacuum field has been explored for force sensing in the optomechanical system \cite{ZHU2024107749}. In a recent experimental demonstration, the entanglement of electro-mechanical modes \cite{Li:2020igh} and two massive mechanical oscillators coupled by a common cavity \cite{Ockeloen-Korppi} had been realized. Moreover, numerous studies have been conducted on the potential to entangle two mechanical oscillators, such as employing coherent laser driving \cite{vitali2007stationary}, via modulated optomechanical systems \cite{chakraborty2018entanglement}, two-tone driving \cite{Woolley14}, with a joint effect of coherent feedback and two-tone driving \cite{Li17}. In addition, several hybrid optomechanical schemes have been reported to enhance the degree of entanglement in the two oscillating mirrors, such as oscillators coupled via Coulomb field force with ensemble of two-level atoms \cite{sohail2020enhancement}, a gain medium with three-level cascaded atoms \cite{bekele2023entanglement} and single-atom Raman laser \cite{teklu2018cavity} in doubly resonant optomechanical cavity. { Furthermore, several investigations have been conducted on the quantum state transfer of the extra-cavity broadband squeezed light field as a form of squeezed reservoir engineering coupled to optomechanical cavity systems with two movable mirrors. Such quantum state transfer entangles two nanomechanical resonators via feeding single-mode squeezed light to a ring cavity \cite{huang2009entangling}, and injecting phase-squeezed along with amplitude-squeezed state laser lights in double cavity scheme \cite{pinard2005entangling}. In addition to these, coupling two-mode squeezed state light to two separated optomechanical cavity systems enables quantum features, such as mechanical entanglement \cite{zhang2003quantum, mazzola2011distributing, sete2014light, yang2015generation} }.  

Nonlinear crystal media have long been known sources of squeezed light. Li {\em et al.}\cite{li2015enhanced} a coupled Kerr medium with an optical parametric amplifier (OPA) to enhance the entanglement of two mechanical oscillators. Furthermore, mechanical entanglement between two mechanical resonators in a cavity optomechanical system with OPA has been realized \cite{jiao2023tripartite, Habtamu23, Sheng2020, Wei24}. More recently, other researchers have analyzed the impact of a non-degenerate parametric down converter accompanied by quantum feedback control in generating quantum correlations \cite{luo2020quantum} in two movable mirrors and two-mode mechanical squeezing \cite{ Wodedo24} in a doubly resonant optomechanical cavity set up. Considering the former investigations, it is found that incorporating a parametric amplifier in an optomechanical cavity is a current research area and an efficient way of quantum state transfer to mechanical resonators. With the above motivation and current interest in entangling mechanical resonators, whether in a doubly resonant optomechanical cavity scheme with the quantum features of intracavity squeezing and squeezed vacuum reservoir transfer to two mechanical resonators. This question is motivated by the recent interest in entanglement generation and multimode investigations of light entanglement enhancement by coupling nonlinear optical cavities.

This work proposes a scheme for optimizing entanglement in nanomechanical resonators through quantum squeezing and parametric amplification within an optomechanical framework. We study a hybrid system of driven doubly resonant optomechanical cavities. Specifically, the system is coupled to a squeezed light source and comprises an embedded pumped non-degenerate optical parametric amplifier (NDOPA). We consider the system operating in a weak coupling regime of the sideband-resolved limit, utilizing red-detuned laser drives and resonant cavity-squeezed fields. These conditions favor cooling mechanical resonators and optimizing the quantum state transfer of the squeezed fields of mechanical resonators. We employ logarithmic negativity \cite{vidal2002computable, Plenio2005} to quantify the continuous-variable (CV) entanglement within the bipartite Gaussian states of the two mechanical modes. The analysis shows that the degree of entanglement between the two movable mirrors is significantly influenced by critical parameters such as the parametric phase and the nonlinear gain of NDOPA, the squeezing strength of the injected squeezed vacuum reservoir, optomechanical cooperativity (controlled by laser drive power), and the mechanical bath temperature (phonon thermal excitation). Furthermore, the system under our investigation is believed to have a potential platform for quantum sensing and information processing. 

The structure of this paper is as follows. In Section ~\ref{sec:2}, the model and its Hamiltonian have been described. In section ~\ref{sec:3}, we elaborate on the system dynamics, including the quantum Langevin equations of mechanical modes in the adiabatic approximation, with the derivation of drift, diffusion, and covariance matrices. In Section ~\ref{sec:4}, formulating entanglement between the nanomechanical resonators with its results and discussions is offered. The conclusion is given in section~\ref{sec:5}.

\section{\label{sec:2} Model and its Hamiltonian}

\begin{figure}[ht!]
\centering
\includegraphics[width=1.1\linewidth]{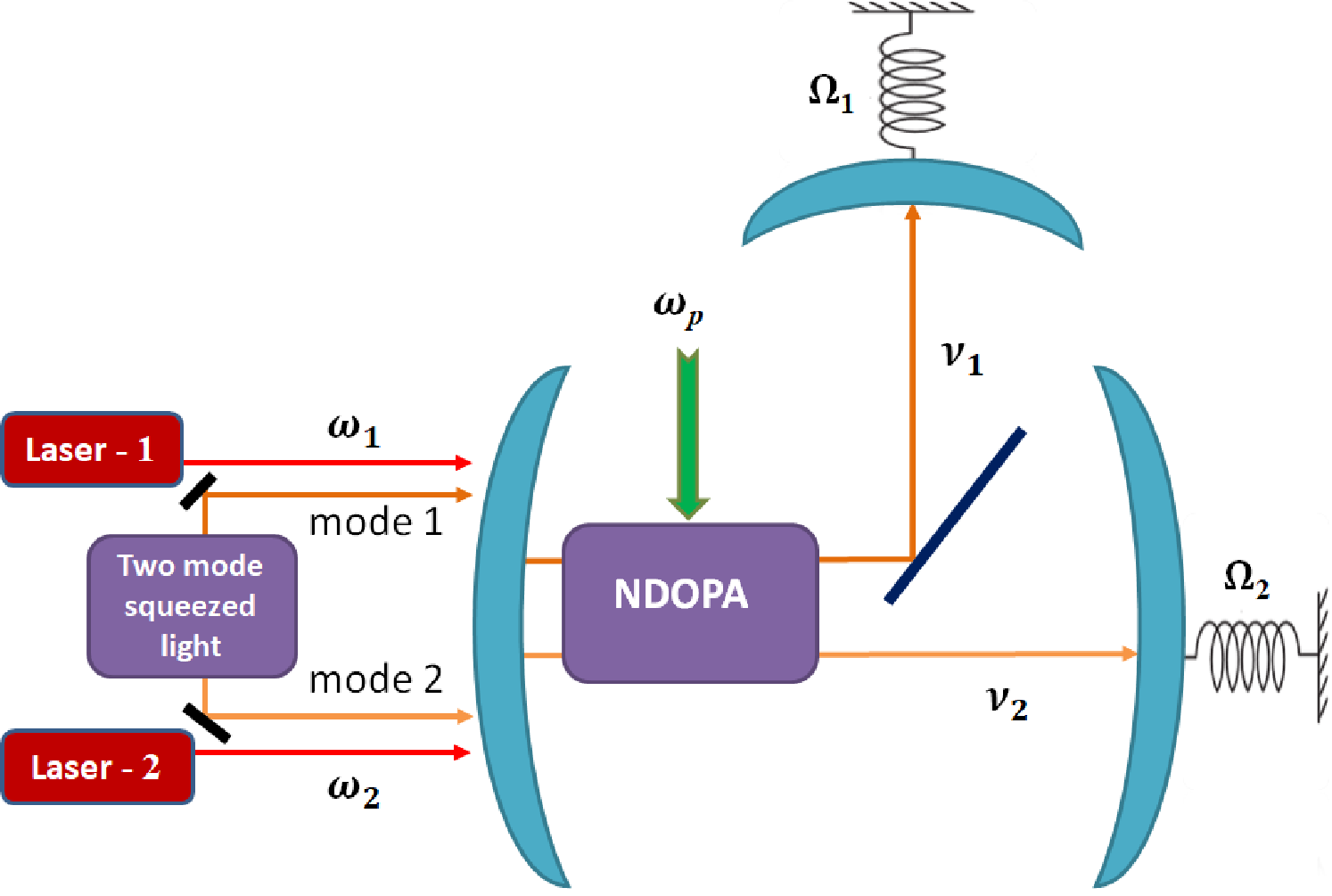}
\caption{\label{fig:1} {Schematic diagram of a doubly resonant cavity system coupled to two nanomechanical resonators of frequencies $\Omega_1$ and $\Omega_2$. It comprises an intracavity NDOPA pumped by a classical laser field frequency $\omega_p$ down-converted into signal and idler photons of frequencies $\nu_1$ and $\nu_2$ respectively, which are resonant with their corresponding cavity mode frequencies. In this case, a two-mode squeezed light source is injected into each cavity through the fixed mirror; meanwhile, each cavity is driven by two coherent lasers of frequencies $\ omega_1$ and $\omega_2$ through the fixed mirror.}}
\end{figure}

We consider a dispersive hybrid optomechanical system as shown in Fig.~\ref{fig:1}, a doubly resonant Fabry-Pérot cavity with a fixed port mirror and two completely reflecting movable mirrors {each oscillating with frequency $\Omega_{j}$. Each cavity field has a resonant frequency of $\nu_{j}$, and a coherent laser of the field of frequency $\omega_{j}$ drives the corresponding cavity through the fixed port mirror.} Moreover, a broadband two-mode squeezed vacuum light is {also coupled to the two cavities through the fixed mirror with a central frequency of $\nu_{j}$}. In addition, the NDOPA is pumped by a coherent laser field {of frequency $\omega_{p}=\nu_1+\nu_2$, is down-converted into correlated signal and idler photon pair of frequencies $\nu_{1}$ and $\nu_{2}$ respectively}. 

The total Hamiltonian $\hat{\mathcal{H}}$ of the system is described by \cite{Schori2, mancini1994quantum}
\begin{align}\label{eq:1}
    \hat{\mathcal{H}}=&\sum ^{2}_{j=1} \hbar\nu_{j} \hat{c}^\dagger_{j}\hat{c}_{j} + \sum ^{2}_{j=1}\hbar\Omega_{j}\hat{d}^\dagger_{j}\hat{d}_{j} -\sum ^{2}_{j=1} \hbar g_{j}\hat{c}^\dagger_{j}\hat{c}_{j}(\hat{d}^\dagger_{j}+\hat{d}_{j}) \nonumber\\ & + 
    \sum ^{2}_{j=1}i\hbar\mathcal{E} _{j}\big(e^{i\phi_{j}}e^{-i\omega_{j}t}\hat{c}^\dagger_{j}-e^{-i\phi_{j}} _{j}e^{i\omega_{j}t}\hat{c}_{j}\big) \nonumber\\ & +
    i\hbar\mu [c_pe^{i\theta}\hat{c}^\dagger_{1}\hat{c}^\dagger_{2}-c^*_pe^{-i\theta}\hat{c}_{1}\hat{c}_{2}].
\end{align} 
Here, the first and the second terms are the sum of the free Hamiltonian of the cavity fields and moving mirrors, respectively. The operators $\hat{c}_{j}(\hat{d}_{j}) $ and $\hat{c}^\dagger_{j}(\hat{d}^\dagger_{j})$ describe the annihilation and creation operators of each cavity (mechanical) that satisfy the commutation relation $ [\hat{\mathcal{O}}_{j},\hat{\mathcal{O}}_{j'}] =[\hat{\mathcal{O}}^\dagger_{j},\hat{\mathcal{O}}^\dagger_{j'}]=0, [\hat{\mathcal{O}}_{j},\hat{\mathcal{O}}^\dagger_{j'}] = \delta_{jj'}$ with $\hat{\mathcal{O}}=\hat{c},\hat{d}$, and $j,j'=1,2$. {The third term is the Hamiltonian of the optomechanical interactions due to radiation pressure with single-photon optomechanical coupling strength $g_{j}=(\nu_{j}/L_{j})\sqrt{\hbar/(2m_{j}\Omega_{j})}$ with $L_{j}$ is the effective initial cavity length and $m_{j}$ is the effective mass of mechanical resonator. The fourth term describes the interaction Hamiltonian due to the coherent laser driving fields and cavity modes coupling with strength $ \mathcal{E}_{j}=\sqrt{\kappa_{j}\mathcal{P}_{j}/({\hbar \omega_{j})}}$ where $\mathcal{P}_{j}$ is the laser driving power and $\kappa_{j}$ is the cavity decay rate, and $\phi_{j}$ is the phase of the coherent driving laser field. The fifth term is the interaction Hamiltonian of the phase-sensitive type-II polarization NDOPA and the cavity modes with coupling strength $\mu$. The NDOPA is pumped by the classical field complex amplitude $c_p=|c_p|\exp(-i\omega_{p}t)$, and $\theta$ is the pumping field phase.}   

{Moreover, the Hamiltonian $\hat{\mathcal{H}}$ with respect to the Hamiltonian $\hat{H}_0=\hbar\omega_{1}\hat{c}^\dagger_{1}\hat{c}_{1}+\hbar\omega_{2}\hat{c}^\dagger_{2}\hat{c}_{2}$ in a rotating frame of unitary transformation operator, $\hat{\mathcal{U}}= \exp({i\hat{H}_{0}t/\hbar})$ makes the laser driving terms time constant. Thus, the newly generated Hamiltonian $\hat{H}=\hat{\mathcal{U}} \hat{\mathcal{H}}  \hat{\mathcal{U}^{\dagger}}-i\hbar\hat{\mathcal{U}}\partial \hat{\mathcal{U}^{\dagger}}/\partial t $ \cite{aspelmeyer2014cavity} takes the following form}  
\begin{align}\label{eq:2}
    \hat{H}=&\sum ^{2}_{j=1} \hbar\Delta'_{j} \hat{c}^\dagger_{j}\hat{c}_{j} + \sum ^{2}_{j=1}\hbar\Omega_{j}\hat{d}^\dagger_{j}\hat{d}_{j}- \sum ^{2}_{j=1}\hbar g_{j}\hat{c}^\dagger_{j}\hat{c}_{j}(\hat{d}^\dagger_{j}+\hat{d}_{j}) \nonumber \\&
+\sum ^{2}_{j=1}i\hbar\mathcal{E}_{j}(e^{i\phi_{j}}\hat{c}^\dagger_{j}-e^{-i\phi_{j}}\hat{c}_{j}) \nonumber \\& +i\hbar\Lambda  \big(\hat{c}^\dagger_{1}\hat{c}^\dagger_{2}e^{i\theta}e^{-i\Delta_{p}t}-\hat{c}_{1}\hat{c}_{2}e^{-i\theta}e^{i\Delta_{p}t} \big) ,
\end{align}
with $\Delta'_{j}=\nu_{j}-\omega_{j}$ and $\Delta_{p}=\omega_{p}-\omega_{1}-\omega_{2}$ are the detuning of cavity mode and NDOPA, respectively, {and $\Lambda=\mu |c_p|$ is the nonlinear parametric gain of NDOPA}. 

\section{\label{sec:3} Dynamics of the system } 

Employing the quantum Langevin equation (QLE) \cite{WallsandMilburn} with the Hamiltonian of Eq.~(\ref{eq:2}), the  dynamics of the system operators $\hat{d}_{j}$ and $\hat{c}_{j}$ take form of
\begin{equation}\label{eq:3}
\begin{split}
\frac{d}{dt}\hat{d}_{j} =& -i\Omega_{j}\hat{d}_{j} + ig_{j}\hat{c}^\dagger_{j}\hat{c}_{j} - \frac{\gamma_{j}}{2}\hat{d}_{j} + \sqrt{\gamma_{j}}\hat{d}^{in}_{j}, \\
\frac{d}{dt}\hat{c}_{1} =& -i\Delta'_{1}\hat{c}_{1} + ig_{1}\hat{c}_{1}(\hat{d}^\dagger_{1} + \hat{d}_{1}) + \mathcal{E}_{1}e^{i\phi_1} \\ 
&+ \Lambda e^{i\theta} e^{-i\Delta_p t} \hat{c}^\dagger_{2} - \frac{\kappa_{1}}{2}\hat{c}_{1} + \sqrt{\kappa_{1}}\hat{c}^{in}_{1}, \\
\frac{d}{dt}\hat{c}_{2} =& -i\Delta'_{2}\hat{c}_{2} + ig_{2}\hat{c}_{2}(\hat{d}^\dagger_{2} + \hat{d}_{2}) + \mathcal{E}_{2}e^{i\phi_2} \\ 
& + \Lambda e^{i\theta} e^{-i\Delta_p t} \hat{c}^\dagger_{1} - \frac{\kappa_{2}}{2}\hat{c}_{2} + \sqrt{\kappa_{2}}\hat{c}^{in}_{2}.
\end{split}
\end{equation}
Here $\gamma_{j}$ is the mechanical damping rate, $\hat{c}^{in}_{j}$ and $\hat{d}^{in}_{j}$ are the annihilation operators related to input noises with zero-mean value associated with each cavity and moving mirror, respectively. The masses of the movable mirrors are on a nano-gram scale and oscillate at high frequencies. The mechanical quality factors $(\mathcal{Q}_{j}=\Omega_{j}/ \gamma_{j}\gg 1)$ are assumed to be very high. In this regard, the noise operator of each mirror can be taken as a Gaussian white noise source. Hence, its evolution can be considered a Markovian process \cite{genes2008robust}. Thus, thermal noise operator $\hat{d}^{in}_{j}$ with nonzero delta correlation functions \cite{vitali2007optomechanical} are:
$\big<\hat{d}^{in,\dagger}_{j}(t)\hat{d}^{in}_{j}(t')\big>=n_{j}\delta(t-t')$ and
$\big<\hat{d}^{in}_{j}(t)\hat{d}^{in,\dagger}_{j}(t')\big>=(n_{j}+1)\delta(t-t')$. Here $n_{j}=\big[ \exp([\hbar\Omega_{j}/(k_B T_{j})\big]-1 )\big]^{-1}$ is the mean thermal excitation (phonon) number in the movable mirror at temperature $T_{j}$ and $k_{B}$ denotes the standard Boltzmann’s constant. On the other hand, the biased noise fluctuation $\hat{c}^{in}_{j}$ of a two-mode squeezed state is injected with central frequency resonant to the cavity mode frequency ($\nu_{j}$). This allows the transfer of optimal entanglement to the mechanical oscillators can be achieved \cite{Jahne9}. The second-order non-zero correlation functions of the noise operator $\hat{c}^{in}_{j}$ are: $\big<\hat{c}^{in,\dagger}_{j}(t)\hat{c}^{in}_{j}(t')\big>=\mathcal{N}_{j}\delta(t-t'),
\big<\hat{c}^{in}_{j}(t)\hat{c}^{in,\dagger}_{j}(t')\big>=(\mathcal{N}_{j}+1)\delta(t-t'), 
\big<\hat{c}^{in}_{j}(t)\hat{c}^{in}_{j'}(t')\big>=\mathcal{M}\delta(t-t')$, and
$\big<\hat{c}^{in,\dagger}_{j}(t)\hat{c}^{in,\dagger}_{j'}(t')\big>=\mathcal{M}^{*}\delta(t-t'); j,j'=1,2   (j \ne j')$ \cite{gardiner2000quantum, yan2015entanglement}. Here $\mathcal{N}_{1}=\mathcal{N}_{2}=\mathcal{N}=\sinh^{2}{(r)}$ and $\mathcal{M}=\exp({i\varphi})\cosh{(r)}\sinh{(r)}$ with $r$ and $\varphi$ are squeezing strength and phase of the squeezing respectively. Here $\mathcal{N}$ is the mean photon number, and $\mathcal{M}$ is the correlation of the two-photon of the two-mode squeezed vacuum reservoir. 
 
According to Eq.~(\ref{eq:3}), the steady-state values of the system operators corresponding to the cavity modes $c^{s}_{j}$ and mechanical modes $d^{s}_{j}$, we have
\begin{equation}\label{eq:4}
\begin{split}
c^{s}_{1}=&\frac{2\mathcal{E}_{1}e^{i\phi_1}(\kappa_{2}-2i\Delta_{2})}{(\kappa_{1}+2i\Delta_{1})(\kappa_{2}-2i\Delta_{2})-4\Lambda^2},\\ 
c^{s}_{2}=&\frac{2\mathcal{E}_{2}e^{i\phi_2}(\kappa_{1}-2i\Delta_{1})}{(\kappa_{2}+2i\Delta_{2})(\kappa_{1}-2i\Delta_{1})-4\Lambda^2},\\
d^{s}_{j}=&\frac{2i g_{j}|c^{s}_{j}|^2 }{\gamma_{j}+2i\Omega_{j}},
\end{split}    
\end{equation}
with $\Delta_{j}=\Delta'_{j}-g_{j}(d^{s*}_{j}+d^{s}_{j})$ is the effective cavity-laser field detuning of each optomechanical cavity caused by radiation pressure. Here we neglect the high-frequency term in the steady-state expression of $c^{s}_{j}$. 

Further, we express each operator of the system as a zero mean quantum fluctuation $\delta\hat{\mathcal{O}}_{j}$ plus the steady state value $\mathcal{O}^{s}_{j}=\big<\hat{\mathcal{O}}^{s}_{j}\big>$ such that $\hat{\mathcal{O}}_{j} = \mathcal{O}^{s}_{j}+\delta\hat{\mathcal{O}}_{j}$. Thus, the fluctuations can be easily found analytically through the linearization approach utilization \cite{mancini1994quantum, WallsandMilburn},  given that the nonlinear effect between the cavity field and the movable mirrors is assumed to be weak. An intense coherent driving laser leads to $|c^{s}_{j}|\gg 1$. Thus, the quantum Langevin equations of Eq.~(\ref{eq:3}) can be safely linearized through neglecting terms with the products quantum fluctuations: $\delta\hat{c}^\dagger_{j}\delta\hat{c}_{j}$, $\delta\hat{c}_{j}\delta\hat{d}^\dagger_{j}$ and $\delta\hat{c}_{j}\delta\hat{d}_{j}$. 
Moreover, introducing slowly varying operators $\delta\tilde{c}_{j}=\delta\hat{c}_{j}\exp({i\Delta_{j} t}), \tilde{c}^{in}_{j}=\hat{c}^{in}_{j}\exp({i\Delta_{j} t})$, $\delta\tilde{d}_{j}=\delta\hat{d}_{j}\exp({i\Omega_{j} t})$ and $\tilde{d}^{in}_{j}=\hat{d}^{in}_{j}\exp({i\Omega_{j} t})$ transform the linearized QLES into 
\begin{equation}\label{eq:5}
\begin{split}
\frac{d}{dt}\delta\tilde{d}_{j}=&    -\frac{\gamma_{j}}{2}\delta\tilde{d}_j
+i\mathcal{G}_{j}(\delta\tilde{c}^\dagger_{j}e^{i(\Delta_{j}+\Omega_{j})t}+\delta\tilde{c}_{j}e^{-i(\Delta_{j}-\Omega_{j})t})
\\ & + \sqrt{\gamma_{j}}\tilde{d}^{in}_{j},    \\
\frac{d}{dt}\delta\tilde{c}_{1}=& -\frac{\kappa_{1}}{2}\delta\tilde{c}_{1}+i\mathcal{G}_{1}(\delta\tilde{d}^\dagger_{1} e^{i(\Delta_{1}+\Omega_{1}) t}+\delta\tilde{d}_{1} e^{i(\Delta_{1}-\Omega_{1}) t})\\ &+ \Lambda e^{i\theta }e^{i(\Delta_{1}+\Delta_{2}-\Delta_{p}) t}\delta\tilde{c}^\dagger_{2} +\sqrt{\kappa_{1}}\tilde{c}^{in}_{1},\\        \frac{d}{dt}\delta\tilde{c}_{2}=& -\frac{\kappa_{2}}{2}\delta\tilde{c}_{2}+i\mathcal{G}_{2}(\delta\tilde{d}^\dagger_{2}e^{i(\Delta_{2}+\Omega_{2}) t}+\delta\tilde{d}_{2} e^{i(\Delta_{2}-\Omega_{2}) t})\\ &+ \Lambda e^{i\theta } e^{i(\Delta_{1}+\Delta_{2}-\Delta_{p}) t}) \delta\tilde{c}^\dagger_{1} 
+\sqrt{\kappa_{2}}\tilde{c}^{in}_{2}.
\end{split}
\end{equation}
Here $\mathcal{G}_{j}=g_{j} |c^{s}_{j}|$ is the effective optomechanical coupling rate, and we take the mean of the cavity field amplitude $c^{s}_{j}$ as a real value via adjusting the phase  $\phi_{j}$ of the laser drive field. 

Moreover, it is assumed that the cavity-driving beam is scattered by the vibrating cavity boundary into two portions \cite{genes2008emergence,genes2008robust}, including the first Stokes $(\omega_{j} - \Omega_{j})$ and anti-Stokes $(\omega_{j} + \Omega_{j})$ sidebands. Subsequently, the optomechanical interaction of the field-mirror entanglement is enhanced. We consider that the system is driven by red-detuned lasers ($\Delta_{j}=\Omega_{j}$), wherein each cavity field is almost resonant with anti-Stokes scattering light. This condition favors cooling of mechanical oscillators and optimizes quantum state transfer  \cite{pinard2005entangling,aspelmeyer2014cavity}. In addition, for high mechanical quality factor ($\Omega_{j}\gg \gamma_{j}$), the mechanical frequency satisfies the condition $\Omega_{j} \gg \mathcal{G}_{j}$, $\Lambda$ as well as the system is operating within the resolved sideband limit $\Omega_{j} \gg \kappa_{j}$. Without loss of generality in this working regime, we can achieve a frequency matching condition of $\Delta_{p}=\Delta_{1}+\Delta_{2}$ via tuning the NDOPA pump at $\omega_{p}=\sum^{2}_{j=1}(\omega_{j}+\Omega_{j})$. Accordingly, we can drop the rapid oscillating terms associated with $\exp({2i\Omega_{j} t})$ in Eq.~(\ref{eq:5}) through rotating wave approximation (RWA).

The dynamics of the system can also be considered adiabatically in the resolved sideband regime for a high mechanical quality factor ($\Omega_{j} \gg \kappa_{j} \gg \gamma_{j})$ and in a very weak coupling limit $(\mathcal{G}_{j} \ll \kappa_{j})$. {For the typical parameters $\gamma_{j}=2\pi\times 140 Hz$, $\kappa_j =2\pi\times215 \times 10^3 Hz$ and $\Omega_j=2\pi\times 947 \times 10^3 Hz $ in the current optomechanical system, the numerical values of the set of parameters $\Lambda< 0.5\kappa_j $ and $\mathcal{G}_{j}\leq 0.1\kappa_j $ well satisfy the inequalities for the assumptions made for resolved sideband regime and weak coupling limit.} Thus, in such conditions, the cavity mode dynamics can follow the mechanical modes, and we can eliminate the cavity mode dynamics such that the photons leave the cavity with a lower interaction with the moving mirrors, allowing an optimal quantum state transfer \cite{pinard2005entangling, Jahne9}. In this regard, we can set in large time approximation that the time rate of each of the cavity modes fluctuation operators sets to zero in the RWA version of Eq.~(\ref{eq:5}). Hence, the linearized form of the QLEs of two nanomechanical resonators becomes
\begin{equation}\label{eq:6}
\begin{split}
\frac{d}{dt}\delta\tilde{d}_{1}&=- \frac{\Upsilon_{1}}{2}\delta\tilde{d}_{1} +\chi\delta\tilde{d}^\dagger_{2} + \sqrt{\Upsilon_{1}} \tilde{D}^{in}_{1},
\\
\frac{d}{dt}\delta\tilde{d}_{2}&=-\frac{\Upsilon_{2}}{2}\delta\tilde{d}_{2} +\chi\delta\tilde{d}^\dagger_{1}+\sqrt{\Upsilon_{2}} \tilde{D}^{in}_{2},
\end{split}
\end{equation}
where $\Upsilon_{1}=\gamma_{1}+\Gamma_{1}$ and $\Upsilon_{2}=\gamma_{2}+\Gamma_{2}$ are the effective damping rates of mechanical modes with $\Gamma_{1}=\mathcal{G}_{1}^2\kappa_{2}/{\mathcal{K}}$ and $\Gamma_{2}=\mathcal{G}_{2}^2\kappa_{1}/{\mathcal{K}}$ are optically-induced damping rates with $\mathcal{K}=\kappa_{1}\kappa_{2}/{4}-{\Lambda}^2$. The factor $\chi=(\mathcal{G}_{1}\mathcal{G}_{2}/{\mathcal{K}})\Lambda \exp({i\theta})$ is the effective coupling strength between the two mechanical modes induced by the driving lasers and NDOPA. The effective  input noise operators to the movable mirrors are $\tilde{D}^{in}_{1}=(a_{1}\sqrt{\kappa_{1}}\tilde{c}^{in}_{1}+b_{1}\sqrt{\kappa_{2}}\tilde{c}^{in,\dagger}_{2}+\sqrt{\gamma_{1}}\tilde{d}^{in}_{1})/\sqrt{\Upsilon_{1}}$ and $\tilde{D}^{in}_{2}=(a_{2}\sqrt{\kappa_{2}}\tilde{c}^{in}_{2}+b_{2}\sqrt{\kappa_{1}}\tilde{c}^{in,\dagger}_{1} +\sqrt{\gamma_{2}}\tilde{d}^{in}_{2})/\sqrt{\Upsilon_{2}} $  with  $a_{1}=i\mathcal{G}_{1}\kappa_{2}/({2\mathcal{K}}) $,
$b_{1}=i\mathcal{G}_{1}{\Lambda}\exp({i\theta})/{\mathcal{K}} $, $a_{2}=i\mathcal{G}_{2}\kappa_{1}/({2\mathcal{K}}) $, $b_{2}=i\mathcal{G}_{2}{\Lambda}\exp({i\theta })/{\mathcal{K}}$. 

In Eq.~(\ref{eq:6}), the dynamics of the two nanomechanical resonators indicate a parametric down-conversion interaction process between the two mechanical modes. In the presence of the NDOPA ($\Lambda \ne 0$) and the cavity driving laser field ($\mathcal{G}_{j}\ne 0$), the dynamics of resonator-1 depends on $\delta\tilde{d}^\dagger_{2}$ and $\tilde{c}^{in,\dagger }_{2}$ meanwhile the dynamics of resonator-2 depends on $\delta\tilde{d}^\dagger_{1}$ and $\tilde{c}^{in,\dagger }_{1}$. Therefore, we can infer that the NDOPA in the system and/or the two-mode squeezed vacuum noise injection can be an agent to generate quantum correlation (i.e., entanglement) between the two spatially separated movable mirrors.

To quantify a continuous-variable entanglement between the nanomechanical resonators, we introduce quadrature operators in Eq.~(\ref{eq:6}). We define quadrature operators as $\delta\tilde {Q}_{j}=(\delta\tilde{d}^\dagger_{j}+\delta\tilde{d}_{j})/\sqrt{2}$ and $\delta\tilde {P}_{j}=i(\delta\tilde{d}^\dagger_{j}-\delta\tilde{d}_{j})/\sqrt{2}$ with their corresponding input noise quadrature operators $\tilde{Q}^{in}_{j}=(\tilde{D}^{in,\dagger}_{j}+\tilde{D}^{in}_{j})/\sqrt{2}$ and $\tilde{P}^{in}_{j}=i(\tilde{D}^{in,\dagger}_{j}-\tilde{D}^{in}_{j})/\sqrt{2}$. Thus, the dynamical equations of the movable mirrors in Eq.~(\ref{eq:6}) can be written as a compact form of a first-order differential equation,
\begin{equation}\label{eq:7} 
\frac{d }{dt}\mathcal{Z}(t)=\mathcal{B}\mathcal{Z}(t)+\mathcal{S}(t). 
\end{equation} 
Here $\mathcal{Z}(t)=\big(\delta{\tilde {Q}}_{1},\delta{\tilde {P}}_{1},\delta{\tilde {Q}}_{2},\delta{\tilde {P}}_{2}\big) ^T$ is the vector encompassing both quadrature operators of the mechanical position and momentum fluctuations, $\mathcal{S}(t) =\big(\sqrt{\Upsilon_{1}}\tilde {Q}^{in}_{1},\sqrt{\Upsilon_{1}}\tilde {P}^{in}_{1},\sqrt{\Upsilon_{2}}\tilde {Q}^{in}_{2},\sqrt{\Upsilon_{2}}\tilde {P}^{in}_{2}\big) ^T$ is the vector of noise contributions from the input quadratures, and $\mathcal{B}$ represents the drift matrix given by 
\begin{equation}\label{eq:8}
\mathcal{B}=\begin{pmatrix}
-\Upsilon_{1}/2 & 0 &|\chi|\cos{\theta} &|\chi|\sin{\theta}\\
0 &-\Upsilon_{1}/2&|\chi|\sin{\theta}&-|\chi|\cos{\theta}\\
|\chi|\cos{\theta}&|\chi|\sin{\theta}&-\Upsilon_{2}/2 &0\\
|\chi|\sin{\theta}& -|\chi|\cos{\theta} &0 &-\Upsilon_{2}/2\\
\end{pmatrix}. 
\end{equation}

Before we quantify entanglement, it is crucial to determine the stable working regime of the system. To study the stability conditions of the system, we consider the steady-state scenario defined by the matrix differential equation Eq.~(\ref{eq:7}). Particularly, the stability condition of the system can be achieved by applying the Routh-Hurwitz stability criterion \cite{Parks1962, dejesus1987routh} that confirms all the eigenvalues of the drift matrix $\mathcal{B}$ in Eq.~(\ref{eq:8}) have negative real parts. {Utilizing the eigenvalue equation $\det(\mathcal{B}-\lambda I)=0$, where $\lambda$ is an eigenvalue of the drift matrix $\mathcal{B}$ and $I$ is a $4\times 4$ unit matrix, we obtain a characteristic polynomial equation of the form $ \lambda^{4}+s_1\lambda^{3} +s_2\lambda^{2}+s_3\lambda+s_4 =0$ with coefficients $s_1=\Upsilon_{1}+\Upsilon_{2}$, $s_2=\big(\Upsilon^2_{1}+\Upsilon^2_{2}+4\Upsilon_{1}\Upsilon_{2}-8|\chi|^2\big)/4$, $s_3=(\Upsilon_{2}+\Upsilon_{2})\big(\Upsilon_{1}\Upsilon_{2}/4-|\chi|^2\big)$ and $s_4=\big(\Upsilon_{1}\Upsilon_{2}/4-|\chi|^2\big)^2$. Thus, employing the Routh-Hurwitz stability criterion, the coefficients $s_j>0$ as well as $s_1s_2-s_3>0$, and $s_1s_2s_3-s^2_3-s^2_1s_4>0$. Consequently, we obtain the following three stability conditions for the system as} 
\begin{equation}\label{eq:9}
    \begin{split}
h_1=&\frac{1}{4}\Upsilon_{1}\Upsilon_{2} -|\chi|^2>0,  \\ 
h_2=&\frac{1}{4} (\Upsilon_{1}+\Upsilon_{2}) \big (\Upsilon_{1}^2 +\Upsilon_{2}^2 + 3\Upsilon_{1}\Upsilon_{2}-4|\chi|^2 \big) > 0,  \\ 
h_3=&(\Upsilon_{1}+\Upsilon_{2})\big[h_2  
-(\Upsilon_{1}+\Upsilon_{2})h_{1}\big]h_{1} >0.
    \end{split}
\end{equation}
{These stability conditions confirm that our system is stable and lies within the stable regime. This ensures that the system operates in a stable regime across all presented results.} We also note that the system stability conditions are independent of the phase $\theta$ of the pumping field of the parametric amplifier. Moreover, for high-quality factor mechanical oscillators with efficient laser cooling, the intrinsic damping rate is much smaller than the optically induced damping rate, i.e., $ \gamma_{j} \ll \Gamma_{j}$ \cite{pinard2005entangling, Jahne9}. Hence, for such a case, the stability conditions of the system in Eq.~(\ref{eq:9}) can be reduced to a more simplified form, $\Lambda  < \sqrt{\kappa_{1}\kappa_{2}}/2$. 

The formal solution of the matrix equation Eq.~(\ref{eq:7}) turns out to be 
\begin{equation}\label{eq:10}
\mathcal{Z}(t)= \mathcal{Z}(0)e^{\mathcal{B}t}+\int_{0}^{t} e^{\mathcal{B}t'}\mathcal{S}(t-t')dt',
\end{equation}
where $\mathcal{Z}(0)$ is the vector of the initial values of the mechanical quadrature components. Since the noises $\tilde{c}^{in}_{j}(t)$ and $\tilde{d}^{in}_{j}(t)$ described by Markovian process as well as $\delta$-correlated like the input noises $\hat{c}_{j}^{in}(t)$ and $\hat{d}_{j}^{in}(t)$ respectively. The stable solution of Eq.~(\ref{eq:7}) is a unique solution that occurs at a steady state and is independent of the initial conditions. Moreover, the quantum noises are zero-mean Gaussian noises, and the dynamics of the fluctuations are in their linearized form. Thus, the steady-state quantum fluctuations of the bipartite mechanical modes become a zero-mean Gaussian state characterized by a $4\times4$ covariance matrix whose elements can be determined through the relation $\mathcal{R}_{ll'} (\infty)=\big[\big<\mathcal{Z}_l(\infty) \mathcal{Z}_{l'}(\infty)\big> +\big<\mathcal{Z}_{l'}(\infty) \mathcal{Z}_l(\infty)\big>\big]/2$, $l,l'=1,2,3,4$. Thus, for the stable system at a steady state, the covariance matrix $\mathcal{R}$ can be evaluated through the Lyapunov equation \cite{vitali2007optomechanical},
\begin{equation}\label{eq:11}
\mathcal{BR}+ \mathcal{RB}^T=-\mathcal{F},
\end{equation}
where $\mathcal{F}$ is the diffusion matrix whose elements can be obtained by employing the noise vector $\mathcal{S}(t)$ and the $\delta$-correlation noise operators in the definition, $\mathcal{F}_{ll'}\delta(t-t')=\big[\big<\mathcal{S}_l(t) \mathcal{S}_{l'}(t')\big>+\big<\mathcal{S}_{l'}(t') \mathcal{S}_l(t)\big>\big]/2$. Thus, we have
\begin{equation} \label{eq:12}
\mathcal{F}=\begin{pmatrix}
\mathcal{F}_{11} & 0 &\mathcal{F}_{13}  & \mathcal{F}_{14}\\
0 & \mathcal{F}_{11}&\mathcal{F}_{14}&-\mathcal{F}_{13}\\ 
\mathcal{F}_{13}& \mathcal{F}_{14} &\mathcal{F}_{33} &0 \\
\mathcal{F}_{14}& -\mathcal{F}_{13} &0  & \mathcal{F}_{33}\\
\end{pmatrix},
\end{equation}
{where the entries of this diffusion matrix are listed in \ref{A}}.

Accordingly, the covariance matrix $\mathcal{R}$ of the mechanical modes takes the form of
\begin{equation}\label{eq:13}
\mathcal{R}=\begin{pmatrix}
\mathcal{R}_{11} & 0 &\mathcal{R}_{13}  & \mathcal{R}_{14}\\
0 & \mathcal{R}_{11}&\mathcal{R}_{14}&-\mathcal{R}_{13}\\ 
\mathcal{R}_{13}& \mathcal{R}_{14} &\mathcal{R}_{33} &0 \\
\mathcal{R}_{14}& -\mathcal{R}_{13} &0  & \mathcal{R}_{33}\\
\end{pmatrix}.
\end{equation}
Utilizing the drift matrix $\mathcal{B}$ of Eq.~(\ref{eq:8}), the diffusion matrix $\mathcal{F}$ of Eq.~(\ref{eq:12}) and Eq.~(\ref{eq:13}) in the Lyapunov equation Eq.~(\ref{eq:11}), { we obtain a matrix equation with four unknowns of the covariance matrix elements as
\begin{eqnarray*}
\begin{pmatrix}
\frac{-\Upsilon_{1}}{2} & |\chi|\cos{\theta} &|\chi|\sin{\theta}  & 0\\
|\chi|\cos{\theta} & \frac{-(\Upsilon_{1}+\Upsilon_{2})}{2}&   0&|\chi|\cos{\theta}\\ 
|\chi|\sin{\theta}& 0 &\frac{-(\Upsilon_{1}+\Upsilon_{2})}{2} &|\chi|\sin{\theta} \\
0& |\chi|\cos{\theta} &|\chi|\sin{\theta}  & \frac{-\Upsilon_{2}}{2}
\end{pmatrix} 
\begin{pmatrix} \mathcal{R}_{11}\\ \mathcal{R}_{13} \\ \mathcal{R}_{14} \\ \mathcal{R}_{33}\\ \end{pmatrix}=\begin{pmatrix} \frac{\mathcal{-F}_{11}}{2}\\ \mathcal{-F}_{13} \\ \mathcal{-F}_{14} \\ \frac{\mathcal{-F}_{33}}{2} \end{pmatrix}.
\end{eqnarray*}}
{Then we carrying out the Cramer's rule \cite{GONG2002} and we easily find the solutions of the analytic expressions of the entries of the covariance matrix  $\mathcal{R}$ as} 
\begin{equation}\label{eq:14}
    \begin{split}
      \mathcal{R}_{11}=&\frac{\det \mathcal{D}_{1}}{\det \mathcal{D}},
\ \ \ \ \ \ \ \ \ \ \ \mathcal{R}_{13}=\frac{\det \mathcal{D}_{2}}{\det \mathcal{D}}, \\
\mathcal{R}_{14}=&\frac{\det \mathcal{D}_{3}}{\det \mathcal{D}}, \ \ \ \ \ \ \ \ \ \ \
R_{33}=\frac{\det \mathcal{D}_{4}}{\det \mathcal{D}}, 
    \end{split}
\end{equation} 
where the matrix $\mathcal{D}$ is 
\begin{eqnarray*} 
\mathcal{D}&=&\begin{pmatrix}
-\frac{\Upsilon_{1}}{2} & |\chi|\cos{\theta} &|\chi|\sin{\theta}  & 0\\
|\chi|\cos{\theta} & -\frac{(\Upsilon_{1}+\Upsilon_{2})}{2}&   0&|\chi|\cos{\theta}\\ 
|\chi|\sin{\theta}& 0 &-\frac{(\Upsilon_{1}+\Upsilon_{2})}{2} &|\chi|\sin{\theta} \\
0& |\chi|\cos{\theta} &|\chi|\sin{\theta}  & -\frac{\Upsilon_{2}}{2}\\
\end{pmatrix}, \\
\end{eqnarray*} 
{and the matrices $\mathcal{D}_1$, $\mathcal{D}_2$, $\mathcal{D}_3$, and $\mathcal{D}_4$ along with their determinant for sample case are found in \ref{B}}.   

\section {\label{sec:4} Quantifying CV entanglement }

A two-mode CV system's entanglement may be quantified by entanglement monotones such as entanglement negativity \cite{vidal2002computable, Plenio2005} and entanglement of formation \cite{Bennett1996, Giedke2003, Marian2008}. Both may be computed starting from the covariance matrix of the system. Entanglement negativity is commonly employed as an entanglement measure for analytical simplicity. It is defined based on the minimum symplectic eigenvalue of the partially transposed covariance matrix of the Gaussian state. The negativity $E_N$ is given by  
\begin{equation}\label{eq:15} 
E_N = \max[0,-\ln(2V_s)], 
\end{equation}
where $V_s$ is the minimum symplectic eigenvalue of the partially transposed Gaussian state under investigation. This state comprises two mechanical modes of $4 \times 4$ covariance matrix that is re-written as 
\begin{equation*} 
 \mathcal{R}=\begin{pmatrix}
\mathcal{R}_1 & \mathcal{R}_{3}\\
\mathcal{R}^T_{3} & \mathcal{R}_2\\
\end{pmatrix}
\end{equation*}
with $\mathcal{R}_1$, $\mathcal{R}_2$ and $\mathcal{R}_3$ represents a $2\times 2$ sub-block matrices of Eq.~(\ref{eq:13}). Thus, we have the smallest symplectic eigenvalue of the form \cite{adesso2004extremal} 
\begin{equation}\label{eq:16}
V_s = \sqrt{\frac{\zeta-\sqrt{\zeta^2-4\det \mathcal{R}}}{2}},
\end{equation}
where $\zeta=\det \mathcal{R}_1+\det \mathcal{R}_2-2\det \mathcal{R}_3$. The expressions $\det \mathcal{R}_1$, $\det \mathcal{R}_2$, $\det \mathcal{R}_3$ and $\det \mathcal{R}$ are the symplectic invariants. The sufficient condition for the Gaussian bipartite state of mechanical modes is said to be entangled if the inequality $E_N > 0$ or $2V_s < 1$ is satisfied.

Accordingly, we utilize the analytical expressions of covariance matrix elements of Eq.~(\ref{eq:14}) in Eq.~(\ref{eq:15}) to quantify the degree of quantum entanglement in nanomechanical resonators. To quantify the quantum entanglement between the mechanical modes in our scheme, we employ feasible parameters based on recent experiments in the optomechanical system \cite{groblacher2009observation} and certain theoretical values satisfying the parametric working regime. For simplicity, we take identical parameters for field-mirror pairs such that the wavelength of the coherent driving lasers $1064nm (\omega_{1}/2\pi=\omega_{2}/2\pi=2.82 \times10^{14}Hz)$, frequency of the mechanical oscillators $\Omega_{1}/2\pi=\Omega_{2}/2\pi= 947\times 10^3Hz$, masses of the movable mirrors $m_1=m_2=145 ng$, the length of the cavities $L_{1}=L_{2}=25 mm$, the cavity decay rates $\kappa_{1}/2\pi=\kappa_{2}/2\pi=\kappa/2\pi=215\times 10^3 Hz$, and the mechanical modes damping rates $\gamma_{1}/2=\gamma_{2}/2\pi=\gamma/2\pi= 140 Hz$. In addition, we consider equal mechanical bath temperatures (phonon excitations), $T_1 =T_2=T (n_1 =n_2=n)$, and the value of phase squeezing $\varphi=0$. Moreover, we utilize equal laser driving powers, $\mathcal{P}_1=\mathcal{P}_2=\mathcal{P}$ that yield identical optomechanical coupling rates $\mathcal{G}_1=\mathcal{G}_2=\mathcal{G}$. Here we define the optomechanical cooperativity $C$ as $C=4\mathcal{G}^2/(\gamma\kappa)$ which is proportional to power $\mathcal{P}$ of the driving laser field.

\begin{figure}[hb!]
\centering
\includegraphics[width=1.1\linewidth]{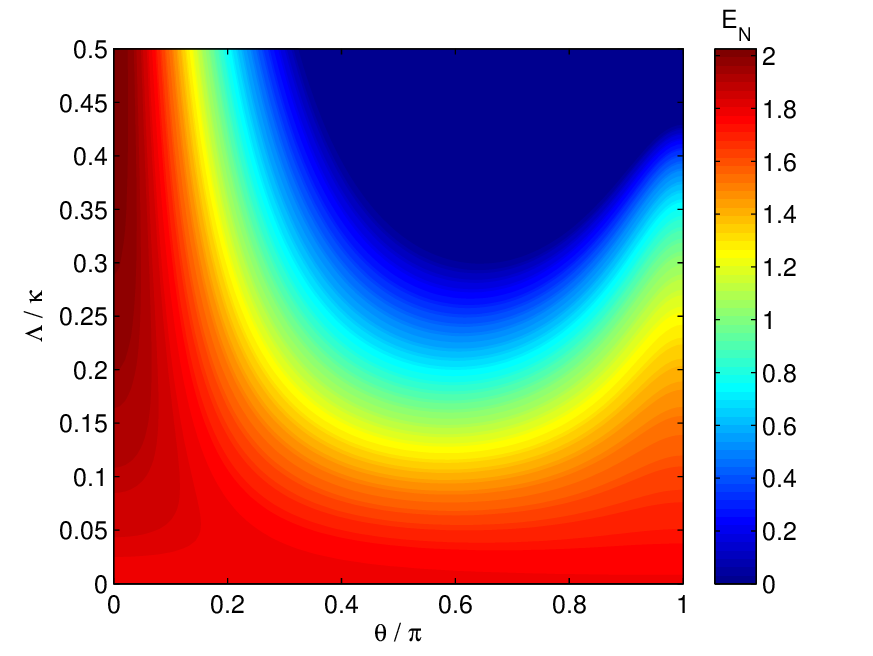}
\caption{Density plot of mechanical entanglement quantification $(E_{N})$ as a function of nonlinear parametric gain versus parametric phase of NDOPA when $r=1$, $T = 42 \mu K (n = 0.5)$, $\mathcal{P}=0.3mW (C=62.5, \mathcal{G}=0.1\kappa)$. The other parameters are listed in the main text.}
\label{fig:2} 
\end{figure}

In Fig.~\ref{fig:2}, the density plot of the entanglement measure as a function of normalized parametric nonlinear gain $(\Lambda/\kappa)$ versus normalized parametric phase ($\theta/\pi$) is demonstrated. In this plot, we observe that increasing the nonlinear parametric gain of NDOPA enhances the degree of entanglement at a parametric phase of $\theta=0$. On the other hand, for higher nonlinear gain, the degree of entanglement decreases from a specific value to zero as the parametric phases increase to $\theta=\pi/2$. In contrast, lower nonlinear gains show relatively higher entanglement and satisfy the CV inseparability criteria for all ranges of parametric phases. Moreover, for lower nonlinear gains, the degree of entanglement decreases as the parametric phases increase towards $\theta=\pi/2$. At the parametric phase $\theta=\pi$, it is illustrated that the degree of entanglement is reduced as the nonlinear gains enhance. 

\begin{figure}[hb!]
\centering
\includegraphics[width=1.1\linewidth]{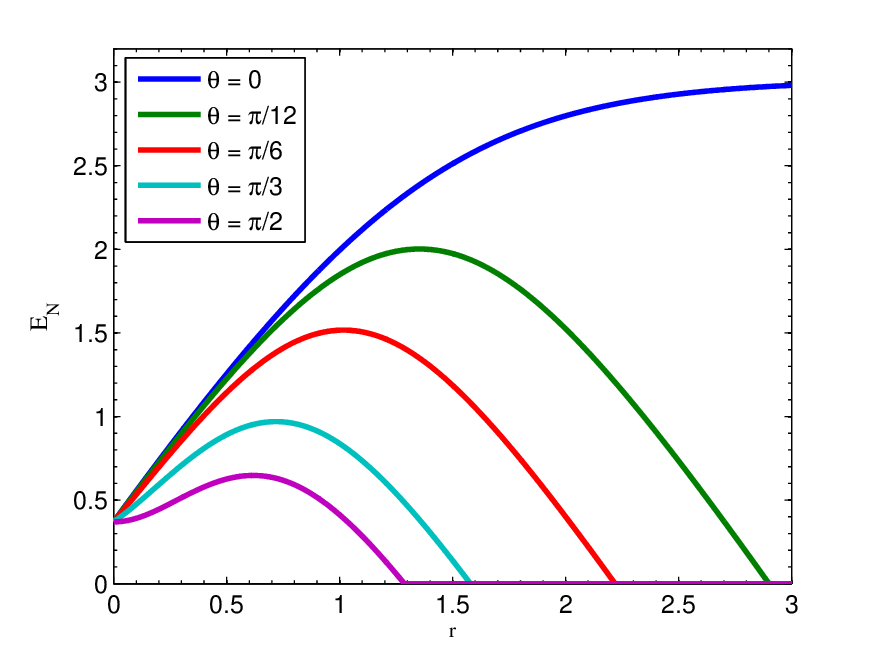}
\caption{Plot of entanglement $(E_{N})$ versus squeezing strength for different parametric phases of NDOPA when the nonlinear parametric gain $\Lambda=0.26\kappa$. The other parameters are the same as in Fig.~\ref{fig:2}.}
\label{fig:3} 
\end{figure}

The entanglement versus the squeezing strength $r$ for different parametric phases are plotted in Fig.~\ref{fig:3}. It is clearly shown that as the parametric phase increases to $\pi/2$, the degree of entanglement decreases. For $r=0$, the entanglement values show no sensitivity for parametric phase variation. This entanglement value is due to the presence of intracavity-squeezed photons of NDOPA. Moreover, corresponding to each parametric phase, as the squeezing strength increases, the degree of entanglement increases to an optimal value and then decreases except at $\theta=0$. It is clearly shown that a broader range of the squeezing strength $r$ of injected squeezed fields corresponds to lower values of parametric phases with enhanced entanglement. For considerable squeezing strength, entanglement in mechanical oscillators becomes a constant for $\theta=0$ or starts to degrade for $\theta \ne 0$ at particular squeezing strength owing to an increase in impurity that comes from the evolution of a pure squeezed state of {photons with $\mathcal{N}=\sinh^{2}{(r)}$ to a thermal squeezed state of photons $\mathcal{N'}>\mathcal{N}=\sinh^{2}{(r)}$ leaving $\mathcal{M}=\cosh{(r)}\sinh{(r)}$ unchanged and thus the phenomenon of a higher decoherence  happened at large value of squeezing strength $r$\cite{Jahne9, Wen-ju2013}. Thus, the quantum state transfer of thermal squeezed photons to nanomechnical resonators affects the mechanical entanglement}.

\begin{figure}[ht!]
\centering
\includegraphics[width=1.07\linewidth]{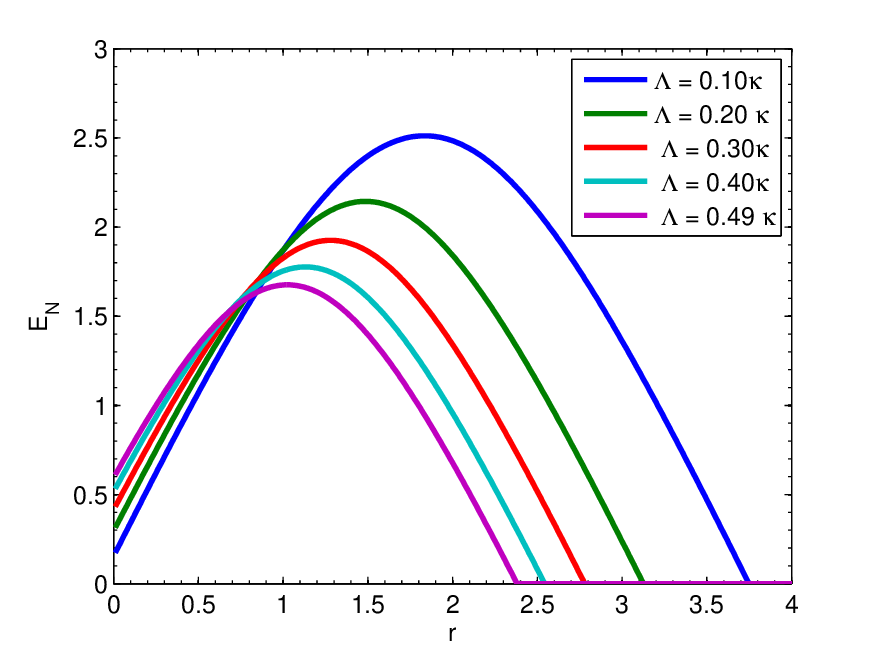}
\caption{Plot of Entanglement $(E_{N})$ versus squeezing strength for different parametric nonlinear gain of NDOPA when $\theta=\pi/12$. The other parameters are the same as in Fig.~\ref{fig:2}.}
\label{fig:4}
\end{figure}

The laser that pumps the NDOPA induces coherent superposition, which is responsible for a strong dependence of the quantum correlation properties of the radiation on the nonlinear gain $ \Lambda $ proportional to the amplitude of the pumping field. In Fig.~\ref{fig:4}, in the absence of the injected squeezed vacuum fields (squeezing $r= 0$) and as the NDOPA nonlinear gain, $ \Lambda $, goes to zero, the mechanical modes entanglement takes the minimal value. In the existence of pumped NDOPA, the generated correlated photons cause entanglement in mechanical modes for $r= 0$, and as the parametric nonlinear gain increases, the entanglement also enhances. Moreover, the joint effect of NDOPA and injected squeezed light makes the entanglement more robust in a certain limit of lower squeezing strength range, $r\ne 0$. The fact that entanglement can explain this is very directly related to the effective couplings between the two mechanical modes, $\chi$ (relies directly on the respective terms, $ \Lambda $), and the correlation matrix element can be increased by squeezing strength $r= 0$ so that the entanglement becomes robust. Furthermore, the effect of NDOPA is more significant for lower squeezing strength, whereas with higher squeezing strengths, the variation of nonlinear gain of NDOPA has negative effects. 
{For considerably higher squeezing strength and nonlinear parametric gain, mechanical entanglement starts to degrade for a similar reason as in Fig.~\ref{fig:3} \cite{Jahne9, Wen-ju2013} }.

\begin{figure}[ht!]
\centering
\includegraphics[width=1.1\linewidth]{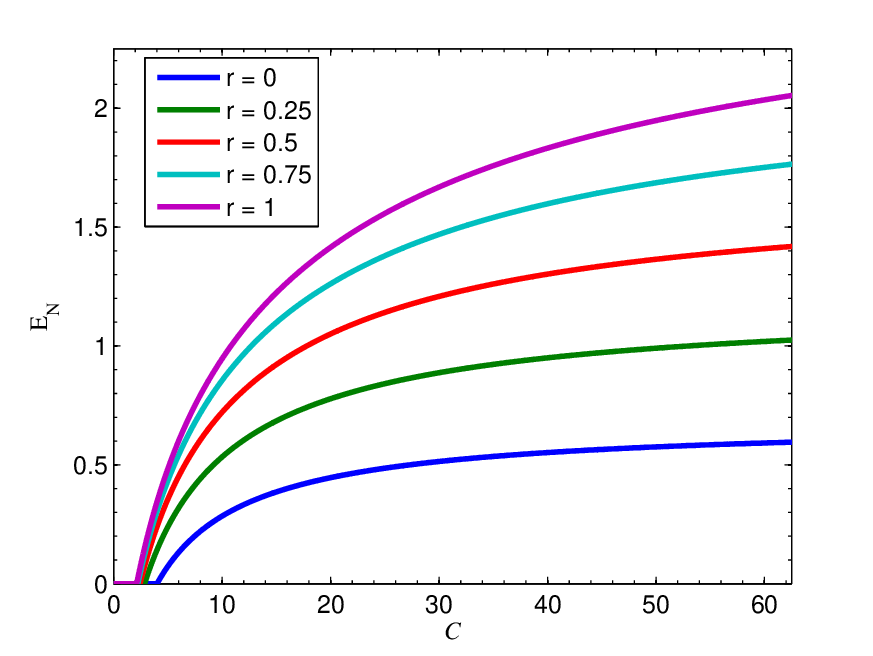}
\caption{Plot of Entanglement $(E_{N})$ versus cooperativity for different squeezing strength when NDOPA parametric phase $\theta=0$ the nonlinear parametric gain $\Lambda=0.49\kappa$. The other parameters are the same as in Fig.~\ref{fig:2}.}
\label{fig:5}
\end{figure}

\begin{figure}[ht!]
\centering
\includegraphics[width=1.1\linewidth]{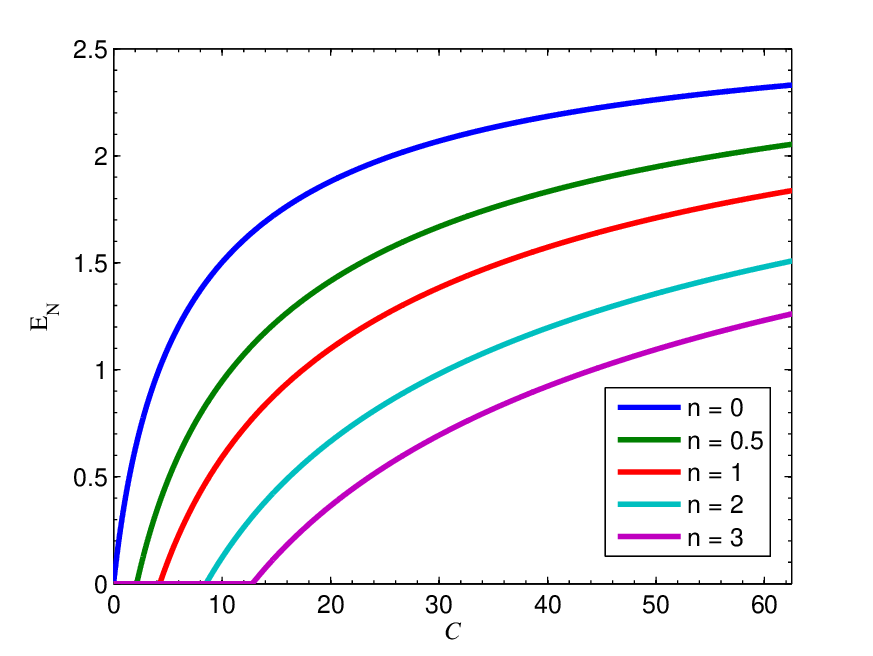}
\caption{Plot of Entanglement $(E_{N})$ versus cooperativity  for different phonon number when NDOPA parametric phase $\theta=0$ the nonlinear parametric gain $\Lambda=0.49\kappa$. The other parameters are the same as in Fig.~\ref{fig:2}.}
\label{fig:6}
\end{figure}

Furthermore, the effects of optomechanical cooperativity $C$ in CV entanglement are plotted in Fig.~\ref{fig:5} and Fig.~\ref{fig:6} for different squeezing strengths $r$ and thermal phonon numbers $n$, respectively. In Fig.~\ref{fig:5} and Fig.~\ref{fig:6}, we demonstrated clearly that as the cooperativity enhanced, the degree of entanglement drastically increased for lower cooperativity values while the degree of entanglement increased slowly for higher cooperativity. It is also shown that the degree of entanglement is enhanced due to the effect of increasing $r$. However, increasing phonon number $n$ is negatively related to the degree of entanglement. Thus, the degradation of mechanical entanglement is due to the elevation of phonon excitations (thermal decoherence effect) in mechanical modes.
Furthermore, the effects of cooperativity can be explained via the degree of entanglement related directly to the square of the strength of the optomechanical coupling $\mathcal{G}^2$ due to the radiation pressure or directly proportional to the laser driving power $\mathcal{P}$ so that the entanglement can be enhanced. In addition, the many-photon coupling $\mathcal{G}$ depends directly on the amplitude of the driving laser, resulting in an enhancement of intracavity photon numbers. In addition, there is a minimum cooperativity requirement to initiate entanglement in the mechanical modes corresponding to each squeezing strength in Fig.~\ref{fig:5} and thermal photon number in Fig.~\ref{fig:6}. In Fig.~\ref{fig:5}, it is shown that the smaller the squeezing strength, the greater the cooperativity requirement for the mechanical modes to display a non-classical effect. Similarly, in Fig.~\ref{fig:6}, the smaller the thermal phonon number corresponds to a lower quantity of cooperativity needed to attain entanglement.


\begin{figure}[ht!]
\centering
\includegraphics[width=1.1\linewidth]{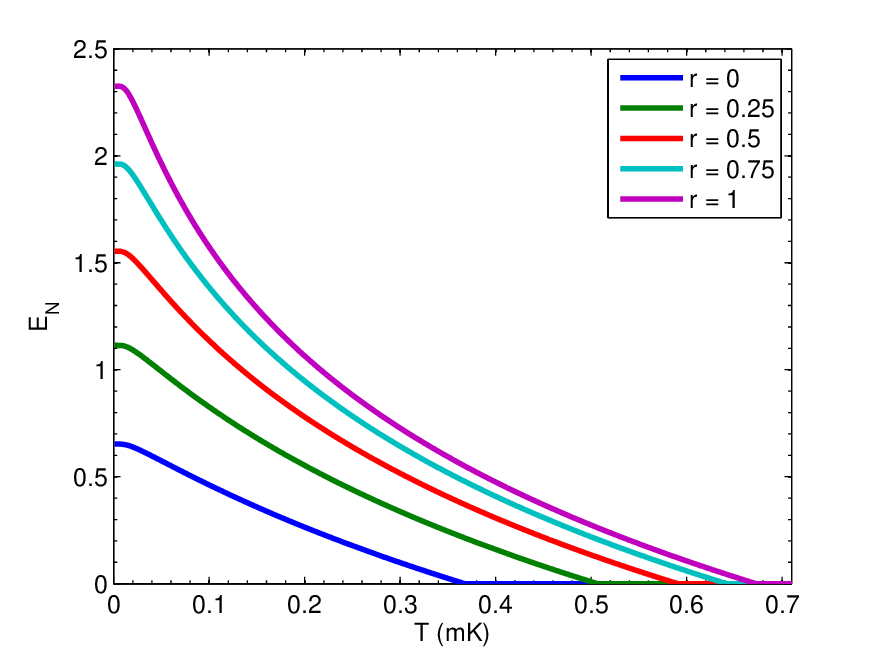}
\caption{Plot of Entanglement $(E_{N})$ versus temperature for different squeezing strength when NDOPA parametric phase $\theta=0$ the nonlinear parametric gain $\Lambda=0.49\kappa$. The other parameters are the same as in Fig.~\ref{fig:2}.}
\label{fig:7}
\end{figure}

It is also essential to know the effects of the temperature in mechanical baths, which is the resource for decoherence and diminishes entanglement. To this end, the plot of entanglement as a function of the temperature $(T )$ of mechanical baths is shown in Figs.~\ref{fig:7} for different squeezing strengths $(r)$ for fixed parametric phase $\theta=0$ and nonlinear gain $\Lambda = 0.49\kappa$. In particular, at $T=0K$ as the nonlinear gain approaches its threshold value for $r=0$ ( absence of squeezed vacuum coupling), the entanglement is limited to a maximum of $ E_N=\ln {2} \approx 0.693$ (equivalent to 3dB) due to the stability condition \cite{luo2020quantum, Wodedo24}. The system has to be coupled with a broadband two-mode squeezed vacuum reservoir $r>0$ to enhance the entanglement above the limit of $E_N=\ln {2}$. Thus, it is also clearly depicted that as the squeezing strength increases, the entanglement becomes more robust against temperature. It can be inferred that the higher mechanical bath temperature increases vibration and enhances phonon numbers directly related to thermal decoherence, significantly degrading the coherence vital for entanglement. Moreover, the temperature of mechanical baths is negatively related to the degree of entanglement at various squeezing strengths. 

In summary, regarding the entanglement generation in the nanomechanical resonators in our scheme, we illustrated logarithmic negativity as a CV entanglement measure depicted in the plots Figs.~\ref{fig:2}-~\ref{fig:7} for the same set of parameters to the corresponding figures. These plots are a function of the parametric pump phase and the nonlinear gain of NDOPA, the strength of the injected squeezed vacuum field, the temperature (phonon excitation) of the mechanical baths, and the optomechanical cooperativity. The recent experimental feasible parameters in an optomechanical system \cite{groblacher2009observation} are relevant for quantifying CV entanglement in mechanical modes. Moreover, one infers the feasibility of this quantitative entanglement found in the working regime of our scheme in Ref.\cite{Wodedo24} by measuring the covariance matrix elements of the output cavity modes through the experimental technique suggested in Ref.\cite{Laurat2005, mazzola2011distributing} through two simultaneous homodyne detections. In the future, we recommend that in the current scheme, it is possible to investigate further quantum correlations in mechanical modes, including Gaussian quantum steering, Gaussian quantum discord, Gaussian interferometric power, Bell nonlocality, and so on. 

\section{\label{sec:5} Conclusions}

We have investigated enhancing entanglement in mechanical resonators coupled through radiation pressure to a doubly resonant cavity incorporating intracavity squeezed fields generated by a non-degenerate optical parametric amplifier and an extra cavity two-mode squeezed vacuum injection. In particular, we consider that the system is operating within a weak coupling and good cavity limit. We employ the covariance matrix entries in logarithmic negativity to quantify the entanglement of continuous-variable Gaussian bipartite states of the mechanical modes. Our findings show that the degree of entanglement is significantly impacted by factors such as the squeezing strength, the phase of the parametric pump, the nonlinear gain of the non-degenerate optical parametric amplifier, optomechanical cooperativity, and the temperature of the mechanical baths. Optimal entanglement occurs at a specific phase choice where noise suppression is maximized. Consequently, increasing the squeezing parameter further makes the entanglement more robust. Additionally, increasing the nonlinear parametric gain can strengthen the entanglement between the resonators, even at lower squeezing strengths. We also observed that coupling a two-mode squeezed vacuum reservoir to the system provides robustness against the thermal noise of the mechanical bath. At higher thermal noise levels, the entanglement can be enhanced by mitigating noise through increased optomechanical cooperativity. These results suggest that the entanglement in the system can be controlled through various parameters, making this model a promising candidate for applications in quantum metrology and quantum information processing.

\section*{Acknowledgments}
BT acknowledges the support provided through project C2PS-8474000137 and project no. 8474000739 (RIG-2024-033). MAW, TYD and TGT  acknowledge ASTU.

\appendix
{\label{app}}

\section{{Diffusion matrix elements}} {\label{A}}
{In this Appendix, there are list of the entries of the diffusion matrix $\mathcal{F}$ of Eq.~(\ref{eq:12})}
\begin{eqnarray*}
\mathcal{F}_{11}&=&\frac{1}{2}\bigg(\big[|a_{1}|^2 \kappa_{1}+| b_{1}|^2 \kappa_{2}\big] (2\mathcal{N}+1)  \\ &&+2\big [b^*_{1}a_{1}\mathcal{M}+a^*_{1}b_{1}\mathcal{M}^*\big]\sqrt{\kappa_{1}\kappa_{2}} +\gamma_{1}(2n_{1}+1)\bigg),  \\ 
\mathcal{F}_{33}&=&\frac{1}{2}\bigg(\big[|a_{2}|^2 \kappa_{2}+| b_{2}|^2 \kappa_{1}\big] (2\mathcal{N}+1)  \\ &&+2\big [b^*_{2}a_{2}\mathcal{M}+a^*_{2}b_{2}\mathcal{M}^*\big] \sqrt{\kappa_{1}\kappa_{2}} +\gamma_{2}(2n_{2}+1)\bigg),     \\ 
\mathcal{F}_{13}&=&\frac{1}{4}\bigg(\big([b^*_{2}a^*_{1} +b_{2}a_{1}] {\kappa_{1}}+[b^*_{1}a^*_{2} +b_{1}a_{2}] \kappa_{2}\big)(2\mathcal{N}+1)  \\ 
&&+ 2\big([b^*_{1}b^*_{2} +a_{1}a_{2}] \mathcal{M}+[b_{1}b_{2} +a^*_{1}a^*_{2}] \mathcal{M}^*\big)\sqrt{\kappa_{1}\kappa_{2}}\bigg),  \\ 
\mathcal{F}_{14}&=&\frac{i}{{4}}\bigg(\big([b^*_{2}a^*_{1} -b_{2}a_{1}] {\kappa_{1}}+[b^*_{1}a^*_{2} -b_{1}a_{2}] \kappa_{2}\big)(2\mathcal{N}+1) \\ 
&&- 2\big([b^*_{1}b^*_{2} +a_{1}a_{2}] \mathcal{M}-[b_{1}b_{2} +a^*_{1}a^*_{2}] \mathcal{M}^*\big)\sqrt{\kappa_{1}\kappa_{2}}\bigg). 
\end{eqnarray*}

\section{{Matrices $\mathcal{D}_1$ to $\mathcal{D}_4$}}{\label{B}}
{Moreover, in this Appendix the matrices $\mathcal{D}_1$ to $\mathcal{D}_4$ found in the expressions of the covariance matrix elements in Eq.~(\ref{eq:14}) are listed in the following manner.}
\begin{eqnarray*} 
\mathcal{D}_1&=&\begin{pmatrix}
-\frac{\mathcal{F}_{11}}{2} & |\chi|\cos{\theta} &|\chi|\sin{\theta}  & 0\\
-\mathcal{F}_{13} & -\frac{(\Upsilon_{1}+\Upsilon_{2})}{2}&   0&|\chi|\cos{\theta}\\ 
-\mathcal{F}_{14}& 0 &-\frac{(\Upsilon_{1}+\Upsilon_{2})}{2} &|\chi|\sin{\theta} \\
-\frac{\mathcal{F}_{33}}{2}& |\chi|\cos{\theta} &|\chi|\sin{\theta}  & -\frac{\Upsilon_{2}}{2}\\
\end{pmatrix},\\
\end{eqnarray*} 
\begin{eqnarray*}
\mathcal{D}_2&=&\begin{pmatrix}
-\frac{\Upsilon_{1}}{2} & -\frac{\mathcal{F}_{11}}{2} &|\chi|\sin{\theta}  & 0\\
|\chi|\cos{\theta} & -\mathcal{F}_{13}&   0&|\chi|\cos{\theta}\\ 
|\chi|\sin{\theta}& -\mathcal{F}_{14} &-\frac{\Upsilon_{1}+\Upsilon_{2}}{2} &|\chi|\sin{\theta} \\
0& -\frac{\mathcal{F}_{33}}{2} &|\chi|\sin{\theta}  & -\frac{\Upsilon_{2}}{2}\\
\end{pmatrix},\\ 
\end{eqnarray*} 
\begin{eqnarray*}
\mathcal{D}_3&=&\begin{pmatrix}
-\frac{\Upsilon_{1}}{2} & |\chi|\cos{\theta} &-\frac{\mathcal{F}_{11}}{2}   & 0\\
|\chi|\cos{\theta} & -\frac{(\Upsilon_{1}+\Upsilon_{2})}{2}&   -\mathcal{F}_{13}&|\chi|\cos{\theta}\\ 
|\chi|\sin{\theta}& 0 &-\mathcal{F}_{14} &|\chi|\sin{\theta} \\
0& |\chi|\cos{\theta} &-\frac{\mathcal{F}_{33}}{2}  & -\frac{\Upsilon_{2}}{2}\\
\end{pmatrix},\\
\end{eqnarray*} 
\begin{eqnarray*}
\mathcal{D}_4&=&\begin{pmatrix}
-\frac{\Upsilon_{1}}{2}& |\chi|\cos{\theta} &|\chi|\sin{\theta}  & -\frac{\mathcal{F}_{11}}{2} \\
|\chi|\cos{\theta} & -\frac{(\Upsilon_{1}+\Upsilon_{2})}{2}&   0&-\mathcal{F}_{13}\\ 
|\chi|\sin{\theta}& 0 &-\frac{(\Upsilon_{1}+\Upsilon_{2})}{2} &-\mathcal{F}_{14} \\
0& |\chi|\cos{\theta} &|\chi|\sin{\theta}  & -\frac{\mathcal{F}_{33}}{2}\\
\end{pmatrix}.\\
\end{eqnarray*}
{We illustrate here the determinants of matrices $\mathcal{D}_1$ to $\mathcal{D}_4$ for a particular case when the NDOPA pumping laser field and the squeezed reservoir phases are $\theta=0$ and $\varphi=0$ respectively. We consider two identical mechanical resonator and cavity parameters ($\gamma_j=\gamma, \kappa_j=\kappa, \mathcal{G}_j=\mathcal{G} $) such that $\Upsilon_{j}=\Upsilon=\gamma+ \mathcal{C}\gamma\kappa^2/(\kappa^2-4\Lambda^2)$ where  $\mathcal{C}=4\mathcal{G}^2/(\gamma \kappa)$ is the optomechanical cooperativity. Thus, the the determinant of the matrices become 
$\det{\mathcal{D}_1}=\Upsilon^3\mathcal{F}_{11}/4+|\chi|\Upsilon^2\mathcal{F}_{13}/2$, $\det{\mathcal{D}_2}=\Upsilon^3\mathcal{F}_{13}/4+|\chi|\Upsilon^2\mathcal{F}_{11}/2$, $\det{\mathcal{D}_3}=(\Upsilon^3-4|\chi|^2\Upsilon)\mathcal{F}_{14}/4$, 
$\det{\mathcal{D}_4}=\Upsilon^3\mathcal{F}_{33}/4+|\chi|\Upsilon^2\mathcal{F}_{13}/2$.
With the aid of the determinant of matrix $\mathcal{D}$, $\det{\mathcal{D}}=\Upsilon^4/4-|\chi|^2\Upsilon^2/2$ and the diffusion matrix elements listed in \ref{A}, the expressions covariance matrix elements of Eq.~(\ref{eq:14}) can be formulated.}

\bibliographystyle{elsarticle-num} 
\bibliography{example}
\biboptions{sort&compress}

\end{document}